\documentclass[aps,prl,groupedaddress]{revtex4-1}
\usepackage{graphicx}

\begin{document}


\title{Ultrafast Bessel beams; advanced tools for laser materials processing}


  \author{Razvan Stoian (*)}
  \author{Manoj K. Bhuyan}
	\affiliation{Laboratoire Hubert Curien, UMR 5516 CNRS, Universit\'{e} de Lyon, Universit\'{e} Jean Monnet, 42000 Saint Etienne, France}
	\email[corresponding author ]{razvan.stoian@univ-st-etienne.fr}
  \author{Guodong Zhang}
  \author{Guanghua Cheng}
	 \affiliation{State Key Laboratory of Transient Optics and Photonics, Xi'an Institute of Optics and Precision Mechanics, CAS, 710119 Xi'an, Shaanxi, China}
  \author{Remi Meyer}
  \author{Francois Courvoisier}
	\affiliation{Institut FEMTO-ST, Universit\'{e} de Franche-Comt\'{e}, UMR 6174 CNRS, 25030 Besan\c{c}on, France}
	


\date{\today}

\begin{abstract}
Ultrafast Bessel beams demonstrate a significant capacity of structuring transparent materials with high degree of accuracy and exceptional aspect ratio. The ability to localize energy on the nanometer scale (bypassing the 100\,nm milestone) makes them ideal tools for advanced laser nanoscale processing on surfaces and in the bulk. This allows to generate and combine micron and nano-sized features into hybrid structures that show novel functionalities. Their high aspect ratio and the accurate location can equally drive an efficient material modification and processing strategy on large dimensions. We review here the main concepts of generating and using Bessel non-diffractive beams and their remarkable features, discuss general characteristics of their interaction with matter in ablation and material modification regimes, and advocate their use for obtaining hybrid micro and nanoscale structures in two and three dimensions performing complex functions. High throughput applications are indicated. The example list ranges from surface nanostructuring and laser cutting to ultrafast laser welding and the fabrication of three dimensional photonic systems embedded in the volume.
\end{abstract}

\maketitle


\section{Introduction}

Nowadays micro/nano- technologies are critically dependent on the development of precise and controllable processing tools able to structure materials with utmost precision and reduced collateral damage. Ultrashort laser processing has thus developed into a performant technology able to take up this challenge, with intrinsic processing capabilities well into the nanoscale \cite{Mis06,Sug14,Sug17,MM07}, approaching and even surpassing the 100\,nm scale \cite{Jog04}. This relies on unmatched abilities to process materials using nonlinear excitation and limited thermal diffusion, generating high-end applications where energy localization in space and time by ultrashort pulses is critical. To optimize structuring in terms of yield, quality, and scale, new concepts of smart laser material processing have emerged, based on the spatio-temporal design of irradiation according to the material response \cite{Stoi03,Stoi10,Bau07,Lu13}. An advanced processing strategy requires in-depth understanding of the irradiation and material transformation process, with the ability to synergetically correlate irradiation and material reaction to energy load. The geometry is equally important to increase processing efficiency and precision, and automated procedures of pulse spatial tailoring emerged and developed \cite{Mau09,Has09,Boo11,Has14,Zh16a}, taking advantage of the geometry of interaction. These involve either shaping geometrically the interaction region in relevant forms, corrected aberration-free beam delivery, parallel processing procedures on user-designed patterns, or three dimensional patterning of transverse intensity profiles along the propagation axis \cite{Mau09,Has14,Zh16a,Cour04,San05,Wil98,Mau08}. Coupled spatio-spectral and spatio-temporal approaches have permitted to downsize and sculpt the focal region for the highest accuracy \cite{He10,Squ10,Boo18}. These take benefit of the large spectral content inherent to a short pulse by spatially and temporally locking the frequencies only at the interaction site. Polarization features can be designed spatially and temporally \cite{Has13,Brix01}. All together, by engineering in space, time, and vectorial domains, \cite{Stoi10,Has14,Amo17,Pad04}, light confinement can harness key features of interaction with matter and build-up unique material transformation paths with highest precision. This is key in designing materials and functions, upscaling approaches for accurate and flexible processing.

Within these developments, recently a new class of ultrafast laser beams emerged for materials processing applications in ultrafast modes \cite{Sch10,BD10}, with potential in attaining processing accuracy beyond diffraction limit, well into the nanoscale domain. These relies on non-diffractive concepts; particularly the Bessel-Gauss beams \cite{McL54,Dur87,Wig91}, where the interaction section can be multidimensionally designed using non-diffractive propagation. The potential of these Bessel-Gauss beams (hereafter called Bessel beams for simplicity) is fully exploitable in transparent materials, with an energy gap superior to the photon energy. Here dielectrics represent a considerable field of action in view of their technological importance in optics, telecom, micro-electronics, or precision mechanics. Application concepts have swiftly emerged in laser structuring, dicing, drilling, in the development of embedded optical elements \cite{Juod01,Juod13,Sug17b,Lu15,Mey17,Ahn18,Yan17,Mit15,Zam09,Wa16,Dud17} in optical materials, or in the precise interaction with biological tissues \cite{Dho07}.

We will review here some of the recent results of ultrafast zero-order Bessel beam applications in material processing, particularly for transparent materials. The benefits of these elongated beams that exceed the Gaussian Rayleigh range by several orders of magnitude for laser micro and nano-processing will be outlined. We will briefly discuss generation concepts, major physical aspects of laser-material interaction in non-diffractive modes and the way to control energy deposition. We indicate several promising applications fields in deep drilling, dielectric cleaving and cutting, materials welding, and the generation of hybrid micro and nanoscale features for embedded photonic devices \cite{BD10,BS14,Rapp17,Zh18,Zh16,Zh18b,Mar17} with applications in spectroscopy and sensing.

\section{Bessel beam interaction with matter}

\subsection{Bessel beam generation}
Several methods for generating non-diffractive Bessel beams (for an extended review see \cite{DA12} and the references therein) rely on conical intersection of wavefronts generated by either a conical phase (axicon, spatial light modulator) and subsequent refraction or diffraction, or by the convergence of annular beams with non-varying polarization. The first approach is among the most utilised as it harnesses most of the input energy of the incident Gauss pulse on the transformation optical element. The phase synchronization and interference of the incoming wavefronts after conical refraction generates an elongated central core of light on distances well beyond of what can be achieved normally by focusing Gauss pulses (i.e. confocal distance), creating the so-called non-diffractive appearance. The core is surrounded by a series of rings given by the higher orders of interference of the conically converging wavefronts. Due to the conical geometry the circular lobes serve as a continuous energy reservoir for the core, feeding its development and favoring a remarkable stability. Depending on the convergence angle and the lateral size of the input beam, the core can develop on significant distances. This axial elongation can be simply seen as a shadow projected on the axis of the initial Gauss transverse dimension (waist) after the conical refraction, or more precisely as the geometrical places where the spatially-limited sweeping front crosses the propagation axis. For practical applications, extremely high aspect ratios can be obtained ($>1000$), where both the elongation ($L$) and the diameter ($d_{FWHM}$) of the core are defined by the conical angle $\theta$ ($L=w_{Gauss}/\tan(\theta)$, $d_{FWHM}\sim 2.4/(k\sin(\theta))$, with $k$ being the wave-vector). The axial distribution can be further modulated using apodizing techniques \cite{Ciz09}. The created beam can be re-imaged, demagnified and steered to the workpiece by relay optics \cite{Bh17b}, with micron-sized cores and aspect ratios exceeds easily a factor 100 in processing conditions. Uniform distributions can be achieved by spectral filtering the leakage from non-ideal axicon tips \cite{Bh17b}. These non-diffractive beams are nonlinearly stable \cite{Gad01,FT12}, robust, as the core is continuously supplied with photons, and less affected by spherical aberrations. This is an essential difference between Gaussian and Bessel illumination that lies in the nonlinear propagation regime. While at ablation-level intensities, Gaussian beams undergo strong Kerr self-focusing and plasma defocusing, resulting in the complex spatio-temporal reshaping, Bessel beams can sustain a more stable propagation-invariant regime \cite{Pol08}. The latter is possible for sufficiently high conical angles, and has also been extended to Bessel vortices \cite{Xie15}.

\begin{figure}[ht]
\begin{center}
\includegraphics[width=11.50cm]{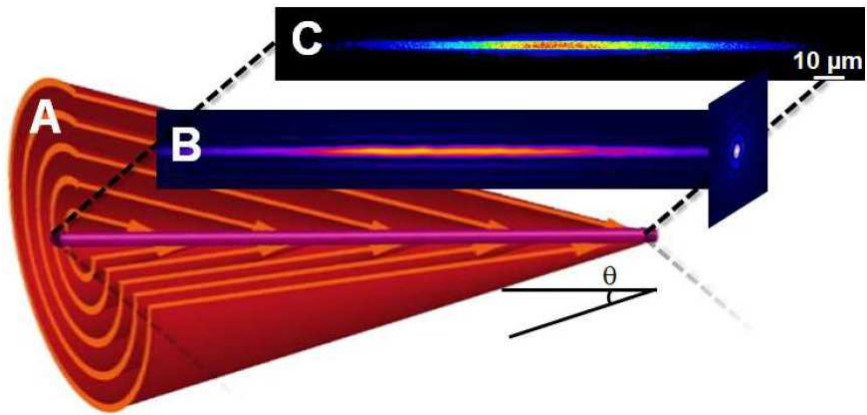}
\caption{Bessel beams. (A) Conceptual scheme of Bessel beam generation. (B) Example of an optically reconstructed core of a Bessel beam with longitudinal and transverse profiles. (C) Plasma in air excited by an ultrashort Bessel beam. \label{Figure1}}
\end{center}
\end{figure}

Examples of Bessel beam generation are represented in Figure~\ref{Figure1}, with a conceptual scheme of conical intersection given in Figure~\ref{Figure1}A, an optically reconstructed beam via sectional transverse imaging (Figure~\ref{Figure1}B) and a plasma trace in air generated by a corresponding Bessel beam (Figure~\ref{Figure1}C), witness of the strong ionization capability of these beams. Most of the discussions below pertain to the wavelength of 800\,nm (Ti:Sapphire laser technology), with few examples at other wavelengths, specified in the text.

The relevant structuring features are related to the dimensional characteristics of the Bessel beams; the high aspect ratio of the intensity distribution with narrow transverse distributions, as well as to its symmetry. Deep and large area structuring can be foreseen. These light longitudinal patterns are versatile enough to allow a distribution of field intensity with a symmetry easily controllable by simple optical transformations \cite{Dud18}. Techniques of programmable amplitude and phase manipulation can exercise further control on the Bessel beams in terms of contrast and uniformity \cite{Ciz09}. The energy ratio between the core and the lobes can equally be engineered (for example by coherently superposing two Bessel distribution via an annular binary aperture), allowing to concentrate more energy on the core, while still retaining a range of non-diffractive elongation \cite{Mo15}. A core-lobe contrast reduction from 16\% to 4\% was demonstrated, trading off the aspect ratio. Recently, concepts of multiplexing Bessel beams have been developed to enhance the structuring throughput \cite{Yu16} with parallel approaches. Some of these shaping possibilities will be indicated below with respect to the relevant application.

\subsection{Interaction of ultrafast Bessel beams with transparent materials}
Due to its interferential nature, the Bessel beam represents a quasi-stationary axial light distribution, where an apparent superluminal front evolution (in case of an axicon) is given by the axial projection of the incoming wavefronts. The ultrashort time envelope and the resulting intensity ensure in a dielectric material carrier excitation via multiphoton processes and collisional multiplication \cite{It06}. The order of the nonlinear absorption is simply defined by the number of photons required to bridge the material forbidden band energy gap. The laser energy is deposited in a free carrier plasma and then relaxes to the host material matrix. Depending on the degree of excitation, the energy can be transferred to the matrix by electronically-induced structural effects intermediated by bond breaking at sub-critical plasma densities, or, if a threshold in the plasma density is achieved, via collisional vibrational activation (electron-phonon coupling), heating, and evolution towards the liquid and the gas phase. Characteristic to the Bessel distribution, the high pressure high temperature state achieved in the material exposed volume relaxes perpendicular to the beam axis. If the internal pressure exceeds the mechanical resistance of the material, shock and rarefaction occurs \cite{ST06}, compressing the material to the sides and opening axially a low density phase on the trace of the Bessel core. This is a main difference with respect to Gauss beams which on surfaces evacuate the ablated material and in the bulk preserve the symmetry of the energy deposition geometry. Depending on the pressure level, the opening may occur in the solid or in the liquid phase. As the size of the modification is defined then by cavitation and material rupture due to stress gradients, the size can be considerably smaller than the beam waist, enabling structuring below diffraction limit and making accessible the nanoscale. Similar sizes will occur also in the event of gas-phase nucleation, provided that the thermodynamic conditions are favorable. These aspects were recently addressed in Refs.~\cite{VS16,RC16,Bh17} for Bessel geometries.

\begin{figure}[ht]
\begin{center}
\includegraphics[width=11.50cm]{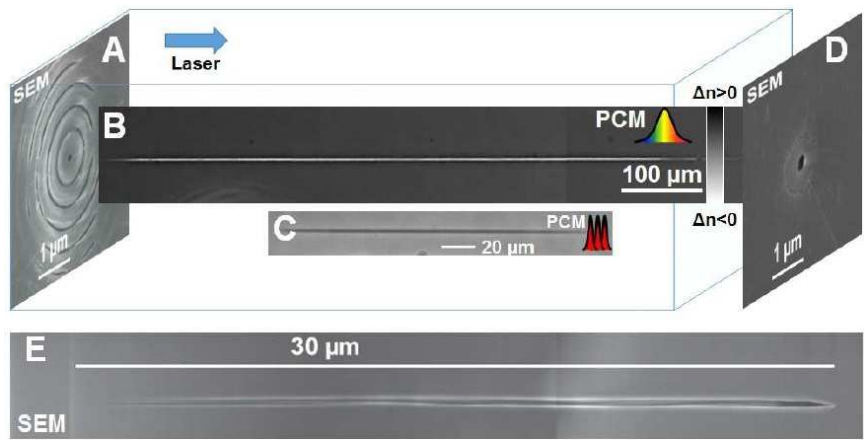}
\caption{Bessel beams interaction with transparent materials \cite{VS16,Bh17}. (A) Input surface pattern generated by a singe shot tightly focused ultrashort (60\,fs, 14\,$\mu$J) Bessel beam ($\theta_{air}=22^{\circ}$) on fused silica; scanning electron microscope (SEM) image. The Bessel ring pattern is recognizable. (B) A high aspect ratio structure in the form of a one-dimensional void in bulk fused silica generated by a single shot ps stretched low focused Bessel beam (2\,ps, 26\,$\mu$J, $\theta_{glass}=4^{\circ}$), viewed as low refractive index domain by phase-contrast microscopy (PCM). (C) Positive refractive index engineering with multishot ($N=500$) ultrashort moderately focused Bessel beam (60\,fs, 1\,$\mu$J, $\theta_{glass}=8^{\circ}$); PCM image \cite{VS16}. Dark colors denote a positive index change and white colors a negative change respectively. (D) Back surface (exit surface) nanoscale structure in fused silica induced by a single shot moderately-focused ps stretched laser pulse (5\,ps, 14\,$\mu$J, $\theta_{glass}=8^{\circ}$); SEM image. (E) Nanoscale void in sapphire (140\,fs, 2\,$\mu$J, $\theta_{sapphire}=15^{\circ}$) \cite{RC16} drilled by a single laser pulse and viewed with SEM after FIB milling. The used energies correspond to fluence values in air (in the absence of nonlinear distortions) from several J/cm$^2$ up to several tens of J/cm$^2$. Laser direction is marked. \label{Figure2}}
\end{center}
\end{figure}

It is first of interest to discuss the Bessel irradiation case of a model glass material such as fused silica (a-SiO$_2$). The types of laser-induced modification range from positive index changes to one-dimensional void generation. Examples \cite{VS16,Bh17} are given in Figure~\ref{Figure2}. The material morphology and the geometry of the modification are controllable via the pulse duration and via the Bessel cone angle (subject to refraction in the bulk). Short pulse durations in the bulk and at low and moderate focusing conditions favor via self-limiting effects soft positive index modification (in the $10^{-4}$--$10^{-3}$ range) \cite{MS13} with interest in developing large core waveguiding elements \cite{DH96}. This is mainly the consequence of a more compact structural rearrangement of matter following bond-breaking \cite{Krol03}. Multiple pulses can determine the spontaneous onset of regular nanostructures \cite{Kaz08,Rud17}. Of interest for material structuring from this work's perspective is a different regime. Stretched laser pulses or tight focusing geometries can overcome diffraction and defocusing and determine strong heating and hydrodynamic movement of the material, opening thus a void. Volume void generation is usually assimilated to micro-explosions \cite{GM97}. Depending on the position of the Bessel beam, structures can be generated on the input surface, in the bulk, or at the exit surface. These cases are presented in Figure~\ref{Figure2}A--E, with examples on single laser shot effects on silica surfaces given in Figure~\ref{Figure2}A, volume structures in silica shown in Figure~\ref{Figure2}B,C, and exit surface patterns depicted in Figure~\ref{Figure2}D. Scanning electron and phase contrast microscopy techniques were used to visualize the structuring results. A further example of bulk void modification in sapphire is given in Figure~\ref{Figure2}E.

Processing the input surface (Figure~\ref{Figure2}A) has the drawback of imprinting the side annular structures, even in the presence of a nonlinear excitation of a higher order. This limits the applicability to energies not significantly exceeding the damage threshold at the core location. Efforts are being made to reduce their impact by further engineering the phase of the Bessel beam \cite{Sug17b}. Equally, multistep irradiation and etching techniques can be applied \cite{Wang18} below the irradiation threshold to enable efficient structuring, where pulse temporal engineering can play a role.

This artefact goes away when considering bulk interactions. The generation of a transient carrier population creates presumably phase instabilities that leads to a dynamic delocalization of the rings, where the optical path difference changes from the condition $\Delta=l \lambda$, with $l>0$. In a general way this suggest indirectly potential beam reshaping due to nonlinearities, cleaning the laser pulse.  This fact can increase the processing window where structures can be made before the lobes can be problematic. Figure~\ref{Figure2}B,C shows bulk interactions for generating voids, respectively refractive index changes in bulk fused silica. The back (exit) surface processing (Figure~\ref{Figure2}D) preserves this self-cleaning of the pulse, exhibiting structures with cross-sections below the micron level without visible side effects. Note that a similar energy was used in Figure~\ref{Figure2}A and Figure~\ref{Figure2}D, many times above the surface threshold, albeit the different pulse duration and focusing conditions ($E_{th}$=0.5\,$\mu$J for tightly focused 60\,fs pulses and $E_{th}$=2.5\,$\mu$J for moderately focused 2\,ps pulses). Sizes even below 100\,nm can be currently achieved \cite{VS16}. This shows a strong potential for processing that will be outlined below. This response is not limited to glass but pertains equally for hard crystalline materials, with an example of a nanovoid in sapphire being shown in Figure~\ref{Figure2}E \cite{RC16}. The technique can be applied to generate through holes in a single shot event, with high reliability and efficiency.

\section{Bessel beam material structuring}
\subsection{Deep surface interaction}
The above considerations lay down a concept of deep processing of transparent materials where the depth of structuring is precisely defined by the position of the Bessel beam and the processing size represents a record in terms of resolution and aspect ratio \cite{BD10}. Surface and bulk interaction processes can be precisely defined, with surface interactions defining a processing window for deep, high aspect ratio drilling as required by specific applications. Micron-sized and sub-micron-sized structures can be patterned, with the depth adjustable by the beam localization with respect to the surface. If input surface interactions present limitations in terms of usable energy range without extended collateral damage, the back surface processing represents a viable strategy for morphing surfaces in depth, minimizing the lobe influence on the damage aspect.

\begin{figure}[ht]
\begin{center}
\includegraphics[width=11.50cm]{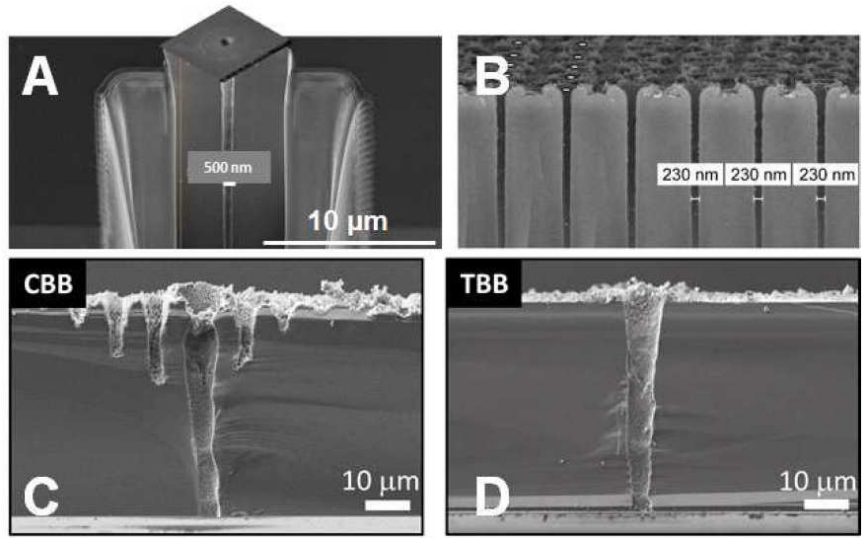}
\caption{Bessel beam surface structuring. (A) Deep structuring of the exit surface in a fused silica parallelepipedic sample with a single shot moderately focused Bessel beam of ps duration (5\,ps, 10\,$\mu$J, $\theta_{glass}=8^{\circ}$). (B) Deep structuring of borosilicate glass \cite{BD10} with single shot tightly focused ultrashort Bessel beams (230\,fs, 0.7\,$\mu$J, $\theta_{glass}=17^{\circ}$). Regular nanostructures of 10\,$\mu$m depth are obtained at the exit surface \cite{BD10}. SEM imaging is used in both cases. (C,D) Deep structures and multi-micrometer-sized through-vias (SEM images) made by nondiffractive beams incident on Si surfaces (courtesy of K. Sugioka (Riken)) \cite{Sug17b} with (C) representing the conventional Bessel beam (CBB) case and (D) a tailored Bessel beam (TBB) for increased core-lobe contrast. A beam at 1.5\,$\mu$m was used. \label{Figure3}}
\end{center}
\end{figure}

Figure~\ref{Figure3}A,B represents an example where two types of glasses are deep-processed on the back surface; Corning 7980-5F fused silica and Corning 0211 borosilicate glass. The result points out the extraordinary precision and aspect ratio achievable with ultrashort and short Bessel beams; a robust pathway to nanostructuring with potential applications. Liquid immersion can be used to remove debris. The scales attainable with this technology are potentially interesting to generate efficient surface functions and optical response in the near and mid-infrared spectral regions. For surface interaction the lobe contrast may pose limitations to the spatial confinement of the damage at high processing fluences due to a first lobe contrast of almost 16\,\%. A method of lobe contrast engineering and Bessel tailoring based on a binary phase was proposed in \cite{Sug17b}, conceptually similar to the Bessel superposition method from \cite{Mo15}, but, due to the phase modulation approach, more energy efficient. An achieved contrast of 0.6\% was reported while keeping the elongation at sufficient values to outpace Gaussian performance. Thus high quality through vias of several tens of microns diameter were demonstrated in silicon layers for three-dimensional integrated circuits using a 1.5\,$\mu$m ultrashort tailored Bessel beam, with photon energies below the gap. The result is illustrated in Figure~\ref{Figure3}C,D. Recently, ultrafast Bessel beams were applied to achieve through-structuring in diamond based on a laser-induced graphitization process all along the irradiation path \cite{Trap17}.

\subsection{Laser cleaving and cutting}
\begin{figure}[ht]
\begin{center}
\includegraphics[width=11.50cm]{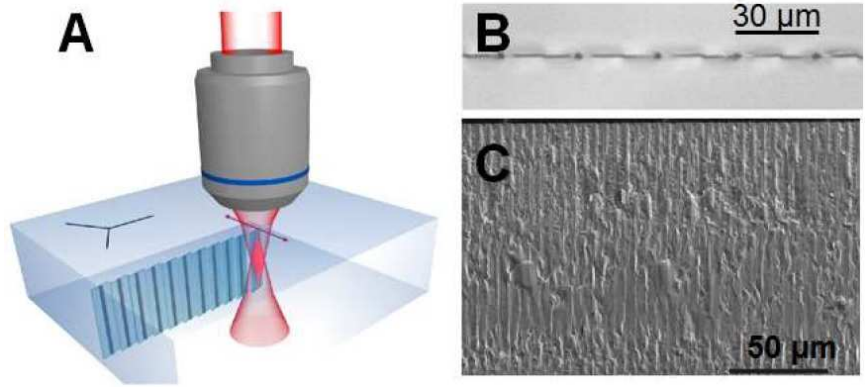}
\caption{Bessel beam cleaving \cite{Mey17,Rapp17}. (A) Experimental scheme indicating the irradiation geometry. For crystalline samples scanning directions with respect to material crystalline orientations are marked on the figure. (B) Directional fracture in c-cut sapphire for cleaving induced with perpendicularly polarized Bessel beams (140\,fs, 20\,$\mu$J, $\theta_{air}=26^{\circ}$) spaced by micron-sized intervals. (C) Cleaved borosilicate glass surface after scanned illumination with elliptical Bessel beams. Ellipticity is introduced in a prior cylindrically symmetric Bessel beam (2.3\,ps, 12\,$\mu$J, $\theta_{glass}=11^{\circ}$). \label{Figure4}}
\end{center}
\end{figure}

The irradiation concept outlined above can be extended to a different application that has a significant technological potential; laser cleaving and cutting of glasses and crystals, where particular care is needed in view of their brittleness. It is to be noted that glass structuring \cite{Itoh14} (and more generally structuring of inorganic transparent materials) represents an important market for displays in consumer electronics or for medical devices as glass shows optical, mechanical and chemical properties of interest. The non-diffractive class of beam is particularly interesting as it can induce large aspect ratio in-volume (ablation-free) modifications of transparent materials for what is called stealth machining \cite{Kuma07} and it can define a promising pathway to glass separation technologies. Figure~\ref{Figure4} gives several examples of laser cleaving of dielectric materials \cite{Mey17,Rapp17} using Bessel beams, with an experimental scheme depicted in Figure~\ref{Figure4}A. The idea behind is to generate a separation path by weakening directionally the bonding strength in the material with the help of an elongated beam. In sapphire (Figure~\ref{Figure4}B), a material of technological importance in displays, high-speed cleaving can be obtained with successive micron spaced Bessel pulses. Ultrashort pulses with controllable polarization direction can induce directional cracks that propagate between impact points, initiating the cleaving process \cite{Rapp17}. The relative orientation to the crystallographic axis and the scan directions are control knobs for the development of the cracks as they define stress fields. Bessel beams allow for creating homogeneous damages all along the depth of the material, where the control on the crack formation enables the processing of a continuous fracture plane, used for the sapphire separation. This opens the route for high speed processing and separation of crystalline substrates. This method, applicable to brittle materials, can combine non-diffractive spatial techniques with temporally-controlled pulse sequences (bursts, tunable envelopes, tunable repetition rates) for a substantial process control \cite{Mish17,Mish16}. This involves tuning the energy deposition for initiating the process while avoiding catastrophic uncontrollable damage or excess heat loading. Recent reports integrated Bessel beam irradiation with additional engineering features (e.g. tilted fronts) to respond to process requirements, for example defining inner contours \cite{Nolte17}. Tilting the axicon can be equally a source of controlling stability and ellipticity, providing an additional element of control \cite{Dud18} in dicing techniques. All these methods can be applied successfully to a large range of crystalline and glassy material by creating directional separation planes.

In isotropic materials, an anisotropy, advantageous for cutting, can be induced by a geometrical transverse deformation of the beam \cite{Mey17,Dud16} affecting the transverse symmetry of the beam along a preferential axis. The associated asymmetric distribution of mechanical constraints will define a plane of separation. An example of a borosilicate glass sample cleaved by elliptical beams \cite{Mey17} is given in Figure~\ref{Figure4}C. The asymmetric beam can successfully perform the cleaving operation based on channel drilling, favoring a preferential axis, without relating to sub-ablation directional stress planes \cite{Mey17}. This represents a novel transverse beam shaping method of non-diffracting beams that can yield a propagation invariant regime at ablation-level intensities. The approach generates elliptical nanochannels in glass after single-shot illumination and the property can be used to guide material cleaving with sub-micron precision.

It is worth noting that the techniques started to acquire a visible industrial dimension, in particular for dicing, cutting, and separation of glassy and crystalline substrates. Patented and patent-pending applications developed at major laser and equipment providers access now a market in full development.

\begin{figure}[ht]
\begin{center}
\includegraphics[width=11.50cm]{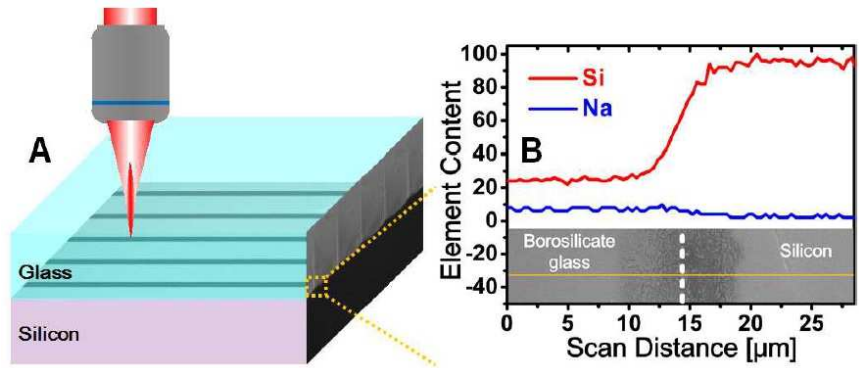}
\caption{Bessel beam welding. (A) Experimental scheme indicating the Bessel beam exposure geometry for glass-silicon welding (140\,fs, 8\,$\mu$J, $\theta_{air}=11.6^{\circ}$).  (B) Elemental analysis at the impact point showing material mixing across the interface. Energy dispersive X-ray spectroscopy was used. A scanning electron microscopy image of the affected zone is provided \cite{Zh18}. \label{Figure5}}
\end{center}
\end{figure}

\subsection{Welding with ultrafast laser Bessel beams}
A straightforward advantage of non-diffractive beams and their longitudinally elongated core on tens and hundreds of microns is the relaxation of the positioning conditions in processing regimes. The energy can be brought to the impact point without precise focusing and positioning \cite{Zh18} approaches. This can be of significant advantage in processing applications, particularly when non-planar surfaces are required or small features (surface nanoprocessing \cite{Cou09,Sah14}), but equally in a different application related to material joining and welding on microscales, where energy has to be precisely transported at the interface \cite{Miy07}. Microwelding occurs point by point or line by line by scanning the laser spot along the contact interface, where nonlinear absorption occurs to heat and change locally the materials.  This requires specifically an accurate laser beam focusing and positioning onto the samples contact interface, a challenge for Gauss beams with micron-sized Rayleigh range. Therefore benefits are expectable for ultrafast Bessel material welding including of dissimilar materials. Figure~\ref{Figure5}A,B shows an example of ultrafast Bessel beam welding of silicon and borosilicate glass (Schott D 263) in optical contact, where the energy is transported across the glass layer to the interface between the materials, with a schematics of the approach shown in Figure~\ref{Figure5}A. The image (SEM) of the affected zone is given in Figure~\ref{Figure5}B together with elemental analysis across the interface in the impact region. Elemental mixing occurs thorough the interface driven by thermal and material flow. The focal-position tolerant zone for efficient welding was reported to increase 5.5-fold by using zero-order Bessel beam processes as compared to Gaussian beam welding \cite{Zh18}. Shear joining strengths as high as 16.5\,MPa were obtained, superior to the Gauss case for similar sizes of processed domains.

We indicate that welding techniques for transparent or dissimilar material are now developed in industrial application by major laser and equipment developers.

\subsection{Hybrid structures for 3D photonics}
\begin{figure}[ht]
\begin{center}
\includegraphics[width=11.50cm]{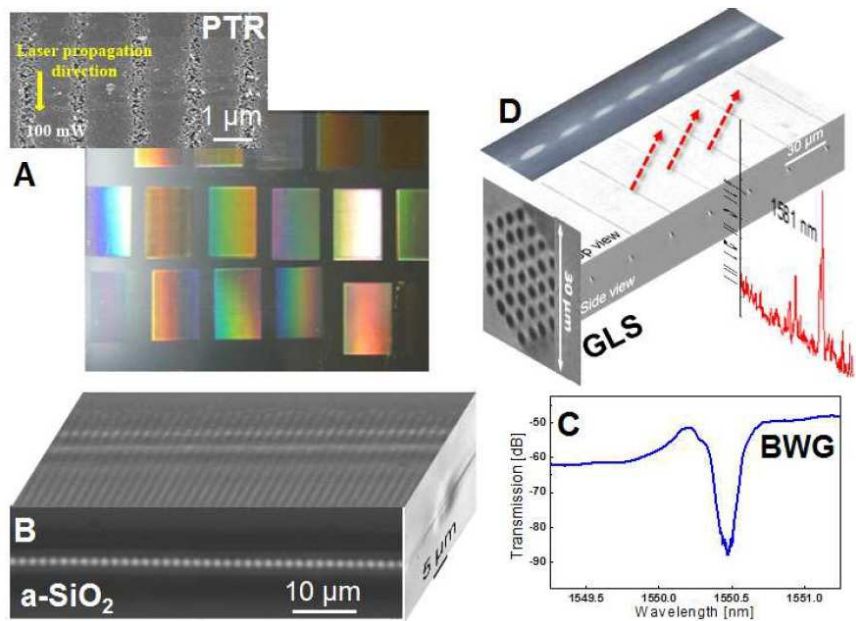}
\caption{Bessel beam fabrication of photonic elements \cite{Zh16,Zh18b,Mar17}. (A) Bessel generated volume gratings in photo-thermo-refractive (PTR) glass and their diffractive color response. Thermal treatment determines crystallite formation in the irradiated zone supporting the refractive index change (inset). (B) Bessel induced 3$^{rd}$ waveguide Bragg gratings in fused silica (single 1\,ps $\mu$J pulses, $\theta_{glass}=15^{\circ}$). (B) Efficient resonant Bragg filtering at 1550\,nm in transmission. (C) Grating sampling the evanescent part of the mode via scattering and diffraction, with a spectrally-resolved response. The grating was fabricated by single ps Bessel pulses ($\theta_{glass}=10^{\circ}$) on the top of a laser fabricated multicore optical waveguide in chalcogenide GLS glass for applications in the near and mid-infrared spectral domains.\label{Figure6}}
\end{center}
\end{figure}

The discussion above has outlined the irradiation conditions for obtaining stable uniform one-dimensional nanovoids or domains of positive index change. These are equivalent to objects of low or high dielectric constant that can be positioned in space. Thus a potential field of interest is related to a high-end application related to 3D photoinscription and development of photonic components using ultrafast lasers, based on concepts of refractive index engineering. The potential for optical functions has been early recognized and optical elements were already tested for fabrication. Putting forward the achievable symmetry, waveguides were photoinscribed using ultrashort Bessel beams \cite{Zam09}. A natural extension of Bessel beam irradiation with respect to the geometrical characteristics, notably the longitudinal elongation, is the fabrication of volume gratings \cite{Juod13} in transparent materials that show high diffraction efficiency. A recent technique that combines Bessel irradiation with thermal treatment was recently reported for high efficacy volume gratings in photo-thermo-refractive glass \cite{Zh16}, where the index changes is delivered via a local crystallization of the glass structure. An example of as-fabricated volume gratings is given in Figure~\ref{Figure6}A, showing diffraction efficiency exceeding 90$\%$ \cite{Zh16}.

Furthermore, a potentially interesting feature of Bessel beam structures comes from their transverse size that can go on the 100\,nm scale \cite{VS16}. Combined with the one dimensional aspect, this feature can be used to sample electrical fields in optical systems, for example in embedded waveguides (e.g. nano-scatterers or nano-diffractors) and create efficient optical functions combining hybrid micro and nanoscale structures. Figure~\ref{Figure6}B,C depicts an example of waveguide Bragg gratings fabricated point by point \cite{Marsh06} by short Bessel pulses in an ultrafast laser photoinscribed waveguide in bulk fused silica \cite{Zh18b}, with its spectral response. A beam at 1030\,nm central wavelength was used to create single pulse sequential nanovoids crossing the section of the guide. The structured waveguide exhibits a highly efficient optical resonance in transmission, of 40\,dB/cm at 1550\,nm in the third order. Ps microjoule laser pulses at 1030\,nm incident wavelength were used to generate uniform nanoscale voids positioned with sub-micron accuracy, enabling also the fabrication of first order gratings and tunable coupling. Bragg resonances in embedded optical waveguides are essential elements for developing mechanically stable optical networks, lasers, and sensing systems for pressure and temperature.

Nanoscale sampling equally allows for the optical reconstruction of light fields with high fidelity by locally scattering field components. The scattering cross-section can be defined by the geometrical section of the Bessel-induced objects and the associated index step. Bessel nanovoids arrays with geometrical sections in the 100\,nm range form thus highly efficient scatterers. Optical field reconstruction is thus of interest and recent applications in astrophotonics were advocated \cite{Mar17}. Figure~\ref{Figure6}D shows an example of an integrated spectroscopy concept working in the telecom range, where the Bessel grating samples and diffracts non-perturbatively the evanescent part of the guided light \cite{Mar17}. The optical system is fabricated in Gallium Lanthanium Sulfide (GLS) chalcogenide glass, making use of a transparency window down to the mid-infrared range. A spectrally-resolved optical signal is generated, characterizing the light source. Accurate optical sampling is at the base of novel integrated guided optics spectrometer concepts based on Fourier or dispersion optical reconstruction of stationary fields with chromatic components to be analyzed \cite{Av06,Co07}.

\section{Conclusions and outlook}
We have reviewed here essential characteristics of ultrafast Bessel laser beams for precise material structuring, notably transparent materials. These beams enable overcoming many of the difficulties usually encountered with standard Gaussian beam focusing in materials and dedicated applications can be defined. Discussing the interaction process, we have outlined their potential to access scales beyond the diffraction limit and their capability in relation to pulse temporal design to controllably transform matter. We have demonstrated via selected examples a range of applicability in surface deep structuring, dielectric cleaving, ultrafast laser welding and generation of novel embedded optical systems in glasses. These rely on high aspect void generation, directional material response, refractive index changes or local heat generation at interfaces. The examples show the applicability of the non-diffractive concepts for deep and large area processing, controlling the material response, as well as for laser cleaving, welding or for developing hybrid ensambles of nano and microscale features for novel photonic functions. At these scales photonic bandgap effects via laser fabrication become feasible.

The field of material processing using laser beams can significantly benefit from advanced optical solutions that are able to master not only the efficiency of the process but equally the dimensional and the structural characteristics of the processed zone with increased levels of control.
In the quest towards novel rapid and reliable processing tools, novel concepts will evolve towards extraordinary classes of light beams enabling interaction in so far unexplored domains and the utilisation of complex light fields such as vortex beams, self-accelerating Airy pulses, or curved beams and spherical caustics is underway \cite{Cour16}.

\section*{Acknowledgements}
  We would like to thank P. K. Velpula, C. D'Amico, L. Rapp, R. Giust, Y. J. Zhang, and G. Martin for their contributions to this work. We acknowledge the Agence Nationale de la Recherche France (projects ANR 2011 BS04010 NanoFlam and ANR 2011 BS09026 SmartLasir) for financial support. We equally acknowledge the support of LABEX MANUTECH-SISE (ANR-10-LABX-0075) of the Universit\'{e} de Lyon, within the program ``Investissements d'Avenir'' (ANR-11-IDEX-0007). This work has been performed in cooperation with the Labex ACTION program, contract ANR-11-LABX-0001-01 and was partly supported by the French RENATECH network.  Research leading to these results has also received funding from the European Union Seventh Framework Programme [ICT 2013.3.2 Photonics] under grant agreement No 619177 TiSa-TD and from the European Research Council (ERC-CoG-682032-PULSAR). We acknowledge the China Scholarship Council, the National Key Research and Development Program ``Laser Manufacturing Technology and Equipment Research on Surfaces of Typical Complicated Components in Aerospace Industry'' (No.2016YFB1102501), and Key Industrial Innovation Chain of Scientific and Technological Innovation Project of Shaanxi Province ``Research and Development of Laser Fine Manufacturing Technology and Equipment for Ultra-hard Materials'' (No.2016KTZDGY-02-02).

\end{document}